# On Identifying Total Effects in the Presence of Latent Variables and Selection bias

**Zhihong Cai**
Department of Biostatistics
School of Public Health
Kyoto University
cai@pbh.med.kyoto-u.ac.jp

**Manabu Kuroki**
Department of Systems Innovation
School of Engineering Science
Osaka University
mkuroki@sigmath.es.osaka-u.ac.jp

## Abstract

Assume that cause-effect relationships between variables can be described as a directed acyclic graph and the corresponding linear structural equation model We consider the identification problem of total effects in the presence of latent variables and selection bias between a treatment variable and a response variable. Pearl and his colleagues provided the back door criterion, the front door criterion (Pearl, 2000) and the conditional instrumental variable method (Brito and Pearl, 2002) as identifiability criteria for total effects in the presence of latent variables, but not in the presence of selection bias. In order to solve this problem, we propose new graphical identifiability criteria for total effects based on the identifiable factor models. The results of this paper are useful to identify total effects in observational studies and provide a new viewpoint to the identification conditions of factor models.

## 1 INTRODUCTION

The evaluation of total effects from observational studies is one of the central aims in many fields of practical science. In observational studies, there may exist latent variables, for example, a variable measured with error, or an unmeasured confounder. On the other hand, observational data may suffer from selection bias, if a sample is selected according to some selection criteria. The existence of latent variables and selection bias hinder the evaluation of total effects from observational data. Many researchers have provided approaches to deal with latent variables in observational studies (Brito and Pearl, 2002; Pearl, 2000; Stanghellini, 2004; Stanghellini and Wermuth, 2005; Tian, 2004). Recently, selection bias has attracted attention from epidemiologists (Greenland, 2003; Hernan et al., 2004), AI researchers (Cooper, 2000) and statisticians (Stanghellini and Wermuth, 2005; Kuroki and Cai, 2006).

In observational studies, it is not rare that both latent variables and selection bias exist in one dataset. However, when we examine current results on latent variables and selection bias, we find that most of them deal with either latent variables or selection bias separately, only a few of them take into account these two situations at the same time. In the presence of latent common causes between a treatment and a response, Pearl and his colleagues provided the back door criterion, the front door criterion (Pearl, 2000) and the conditional instrumental variable (IV) method (Brito and Pearl, 2002) as identifiability criteria for total effects in the framework of linear structural equation models. In addition, in the framework of nonparametric structural equation models, Shpitser and Pearl (2006) and Huang and Valtorta (2006) solved the identification problems of causal effects and provided the complete algorithms to derive the causal effects in the presence of latent variables. In general, these criteria are based on the idea that some observed variables which have no direct association with the latent common causes are used to evaluate the total effects. However, in many practical studies, such latent variables may have an effect on some important observed variables to be used to identify total effects. Under such situations, it is difficult to apply these identification criteria to evaluate the total effects.

On the other hand, when both confounding bias and selection bias may be at work, Spirtes et al. (1999) described the FCI algorithm (Spirtes et al., 2000) as a method to test whether there is a causal path from one variable to another. In addition, Richardson and Spirtes (2002) introduced ancestral graph models as a graphical model in the presence of latent variables and selection bias, and clarified some properties regarding the ancestral graph models. However, these studies fo-cused on the specification problem of causal structure, but not on the identification problem of total effects.

In this paper, we assume that cause-effect relationships between variables can be described as a directed acyclic graph and the corresponding linear structural equation model Then, we consider the problem of identifying total effects from observational studies with latent variables and selection bias between a treatment variable and a response variable Based on the theory of the identifiable factor model, we propose new graphical identifiability criteria to identify total effects under situations where it is difficult to use the identifiability criteria provided by Pearl and his colleagues and Huang and Valtorta (2006). Different from the identification problem of the factor models, it should be noted that we are interested in evaluating the total effects but not the whole causal model. That is, it will be shown in section 3 that there are some situations where the total effect is identifiable even when the whole causal model is not identifiable. These new criteria are useful to identify total effects in observational studies, and they also provide a new viewpoint to the identification conditions of factor models.

## 2 PRELIMINARIES

In statistical causal analysis, a directed acyclic graph that represents cause-effect relationships is called a path diagram. A directed graph is a pair $G = (\boldsymbol{V}, \boldsymbol{E})$, where $\boldsymbol{V}$ is a finite set of vertices and the set $\boldsymbol{E}$ of arrows is a subset of the set $\boldsymbol{V} \times \boldsymbol{V}$ of ordered pairs of distinct vertices. For graph theoretic terminology used in this paper, see, for example, Lauritzen (1996).

Suppose a directed acyclic graph $G = (\boldsymbol{V}, \boldsymbol{E})$ with a set $\boldsymbol{V} = \{V_1, V_2, \cdots, V_n\}$ of variables is given. The graph $G$ is called a path diagram, when each child-parent family in the graph $G$ represents a linear structural equation model

$$V_i = \sum_{V_j \in \text{pa}(V_i)} \alpha_{v_i v_j} V_j + \epsilon_{v_i} \qquad i = 1, \ldots, n, \quad (1)$$

where pa $(V_i)$ is a set of parents of $V_i$. In this paper, if there is no special statement, $\epsilon_{v_1}, \ldots, \epsilon_{v_n}$ are assumed to be independent and normally distributed with mean 0. In addition, $\alpha_{v_i v_j} (\neq 0)$ is called a path coefficient.

The conditional independence induced from a set of equations (1) can be obtained from the graph $G$ according to the d-separation (Pearl, 2000), that is, when $\boldsymbol{Z}$ d-separates $X$ from $Y$ in a path diagram $G$, $X$ is conditionally independent of $Y$ given $\boldsymbol{Z}$ in the corresponding linear structural equation model (e.g. Spirtes et al., 2000). In this paper, it is assumed that a path diagram $G$ and the corresponding joint distribution are faithful to each other; that is, the conditional independence relationship in the joint distribution is also reflected in $G$, and vice versa (Spirtes et al., 2000).

Here, we denote some notations for further discussion. Let $\sigma_{xy \cdot zs^*} = \text{cov}(X, Y | \boldsymbol{Z} = \boldsymbol{z}, \boldsymbol{a} \leq \boldsymbol{S} \leq \boldsymbol{b})$ and $\sigma_{yy \cdot zs^*} = \text{var}(Y | \boldsymbol{Z} = \boldsymbol{z}, \boldsymbol{a} \leq \boldsymbol{S} \leq \boldsymbol{b})$ and $\beta_{yx \cdot zs^*} = \sigma_{xy \cdot zs^*} / \sigma_{xx \cdot zs^*}$ ($\boldsymbol{s}^*$ indicates that each element of $\boldsymbol{s}$ is conditioned by the interval). For disjoint sets $\boldsymbol{X}$, $\boldsymbol{Y}$, $\boldsymbol{Z}$ and $\boldsymbol{S}$, let $\Sigma_{xy \cdot zs^*}$ be a conditional covariance matrix of $\boldsymbol{X}$ and $\boldsymbol{Y}$ given $\boldsymbol{Z} = \boldsymbol{z}$ and $\boldsymbol{a} \leq \boldsymbol{S} \leq \boldsymbol{b}$. We use the same notations in the case where either $\boldsymbol{X}$ or $\boldsymbol{Y}$ is univariate. In addition, let $\Sigma_{yy \cdot xs^*}$ be a conditional covariance matrix of $\boldsymbol{Y}$ given $\boldsymbol{X} = \boldsymbol{x}$ and $\boldsymbol{a} \leq \boldsymbol{S} \leq \boldsymbol{b}$. When $\boldsymbol{S}$ or $\boldsymbol{Z}$ is an empty set, they are omitted from these arguments. Furthermore, let $B_{yx \cdot z} = \Sigma_{yx \cdot z} \Sigma_{xx \cdot z}^{-1}$ be the regression coefficient matrix of $\boldsymbol{x}$ in the regression model of $\boldsymbol{Y}$ on $\boldsymbol{x} \cup \boldsymbol{z}$. The similar notations are used for other parameters.

A total effect $\tau_{yx}$ of $X$ on $Y$ is defined as the total sum of the products of the path coefficients on the sequence of arrows along all directed paths from $X$ to $Y$. In this paper, it is assumed that the readers are familiar with the identifiability criteria for total effects, for example, the back door criterion, the front door criterion (Pearl, 2000) and the IV method (Bowden and Turkington, 1984; Brito and Pearl, 2000). When a total effect can be determined uniquely from the covariance parameters of observed variables, it is said to be identifiable, that is, it can be estimated consistently.

When $\boldsymbol{Z}$ d-separates $X$ from $Y$ in a path diagram $G$, then both $\sigma_{xy \cdot z} = \beta_{yx \cdot z} = 0$ and $B_{yz \cdot x} = B_{yz}$ hold true (e.g. Spirtes et al., 2000).

## 3 IDENTIFICATION OF TOTAL EFFECTS

### 3.1 LEMMA

To derive new graphical identifiability criteria for total effects, we first introduce the following lemmas:

**LEMMA 1**

When $\{X, Y\} \cup \boldsymbol{S} \cup \boldsymbol{T}$ are normally distributed,
$$\beta_{yx \cdot s} = \beta_{yx \cdot st} + B_{yt \cdot xs} B_{tx \cdot s}, \quad (2)$$
$$\sigma_{yy \cdot xs} = \sigma_{yy \cdot x} - B_{ys \cdot x} \Sigma_{ss \cdot x} B'_{ys \cdot x}. \quad (3)$$
□

Equations (2) and (3) are the results of Cochran (1938) and Whittaker (1990), respectively. In addition, the following lemma is given by Wermuth (1989).

**LEMMA 2**

When $\{X, Y\} \cup \boldsymbol{S} \cup \boldsymbol{T}$ are normally distributed, if $\boldsymbol{T}$ is conditionally independent of $X$ given $\boldsymbol{S}$ or $Y$ is condi-tionally independent of $\boldsymbol{T}$ given $\{X\}\cup\boldsymbol{S}$, then $\beta_{yx\cdot st} = \beta_{yx\cdot s}$ holds true. In addition, if $\boldsymbol{T}$ is conditionally independent of $Y$ given $\boldsymbol{S}\cup\{X\}$, then $\sigma_{yy\cdot xst} = \sigma_{yy\cdot xs}$ holds true. □

## 3.2 DUALITY BETWEEN LATENT VARIABLES AND SELECTION BIAS

In this section, we consider two different situations: one is a situation where a latent variable exists shown in Fig.1 (a); the other is a situation where selection bias exists shown in Fig.1 (b), where $\boldsymbol{X} = (X_1, \cdots, X_p)$ is a set of observed variables, and $U$ is a latent variable which has an effect on $\boldsymbol{X}$. In addition, Fig.1(b) indicates that the data have been observed according to the selection criterion $a \leq S \leq b$ (both $a$ and $b$ are possible values of $S$).

Regarding Fig.1 (a), the corresponding linear structural equation model can be provided as

$$X_i = \alpha_{x_i u} U + \epsilon_{x_i}, \qquad i = 1, \ldots, p, \qquad (4)$$

which is called a single factor model (with correlated errors). Then, the covariance matrix of $\boldsymbol{X}$ can be provided as

$$\begin{aligned}\Sigma_{xx} &= \Sigma_{xx\cdot u} + B_{xu} B'_{xu} \sigma_{uu} \\ &= \Sigma_{xx\cdot u} + \Omega(u)\Omega(u)',\end{aligned} \qquad (5)$$

where $\Omega(u)$ is a $p$ dimensional vector, which is called a factor loading in this paper. Here, it should be noted that $\Sigma_{xx}$ can be observed but $\sigma_{uu}$, $\Sigma_{xx\cdot u}$ or $\Sigma_{xu}$ can not be observed.

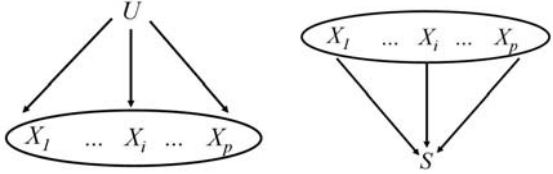

(a): Latent Variable Case   (b): Selection Bias Case

Fig.1: Graphical representation

On the other hand, regarding the $\Sigma_{xx}^{-1}$, by the Sherman-Morrison-Woodbury formula for matrix inversion (Rao, 1972), the inverse matrix of $\Sigma_{xx}$ can be represented as the form of

$$\Sigma_{xx}^{-1} = \Sigma_{xx\cdot u}^{-1} - \Lambda(u)\Lambda(u)', \qquad (6)$$

where $\Lambda(u)$ is a $p$ dimensional vector.

Regarding Fig.1 (b), the corresponding linear structural equation model can be provided as

$$S = \sum_{j=1}^{p} \alpha_{s_i x_j} X_j + \epsilon_{s_i}. \qquad (7)$$

Then, the covariance matrix of the selected population can be provided as

$$\begin{aligned}\Sigma_{xx\cdot s^*} &= \Sigma_{xx} - B_{xs} B'_{xs}(\sigma_{ss} - \sigma_{s^* s^*}) \\ &= \Sigma_{xx} - \Omega(s)\Omega(s)',\end{aligned} \qquad (8)$$

where $\Omega(s)$ is a $p$ dimensional vector (Johnson and Kotz, 1972). Here, $\check{\sigma}_{ss} = \sigma_{ss} - \sigma_{ss\cdot s^*} \geq 0$ since $\sigma_{ss\cdot s^*} = \mathrm{var}(S|a \leq S < \leq b)$ is the variance of a doubly-truncated normal distribution. In the selected population, it should be noted that $\Sigma_{xx\cdot s^*}$ can be observed but $\Sigma_{xx}$, $\check{\sigma}_{ss}$ or $B_{xs}$ can not be observed.

On the other hand, regarding $\Sigma_{xx\cdot s^*}^{-1}$, we can obtain

$$\Sigma_{xx\cdot s}^{-1} = \Sigma_{xx}^{-1} + \Lambda(s)\Lambda(s)', \qquad (9)$$

where $\Lambda(s)$ is a $p$ dimensional vector, which is also called a factor loading in this paper.

In addition, when we discuss latent variable problems and selection bias problems based on conditional distribution given $\boldsymbol{Z}$, the following equations hold true:

$$\Sigma_{xx\cdot z} = \Sigma_{xx\cdot uz} + B_{xu\cdot z} B'_{xu\cdot z} \sigma_{uu\cdot z}$$

and

$$\Sigma_{xx\cdot zs^*} = \Sigma_{xx\cdot z} - B_{xs\cdot z} B'_{xs\cdot z} \check{\sigma}_{ss\cdot z},$$

where $\check{\sigma}_{ss\cdot z} = \sigma_{ss\cdot z} - \sigma_{ss\cdot zs^*} \geq 0$.

From these equations, we can understand that equations (5) and (6) take the same form as equations (9) and (8), respectively. In this paper, such relationships are called the duality between latent variables and selection bias. By using the dual relationships, we will show below that the identification conditions of factor models are useful to solve the selection bias problems.

Let $G_{cov}^{x\cdot u}$ be the undirected graph obtained by connecting any two variables $X_i$ and $X_j$ ($i \neq j$) in $\boldsymbol{X}$ by an undirected edge only if the conditional covariance of $X_i$ and $X_j$ given $U$ is not equal to zero. Let $G_{con}^{x\cdot u}$ be the undirected graph obtained by connecting any two variables $X_i$ and $X_j$ ($i \neq j$) in $\boldsymbol{X}$ by an undirected edge only if the conditional covariance of $X_i$ and $X_j$ given $\{U\} \cup \boldsymbol{X} \backslash \{X_i, X_j\}$ is not equal to zero. When we are concern with the covariance structure of $\boldsymbol{X}$ not conditioning on $U$, $U$ are omitted from these arguments.

Then, Stanghellini and Wermuth (2005) provided the following lemma.

### LEMMA 3

Equations (5)( or equation (6)) can be solved with respect to $\Sigma_{xx\cdot u}$ and $\Omega(u)\Omega(u)'$ (or $\Sigma_{xx\cdot u}^{-1}$ and $\Lambda(u)\Lambda(u)'$) if and only if one of the following conditions holds true:

(1) $\Omega(u) \neq \boldsymbol{0}$ and the structure of zeros in $\Sigma_{xx\cdot u}$ is such that every connectivity component of the complementary graph of $G_{cov}^{x\cdot u}$ contains an odd cycle;

(2) $\Lambda(u) \neq \mathbf{0}$ and the structure of zeros in $\Sigma_{xx \cdot u}^{-1}$ is such that every connectivity component of the complementary graph of $G_{con}^{x \cdot u}$ contains an odd cycle. □

The similar results hold true for equations (8) and (9)

## 3.3 IDENTIFIABILITY CRITERION: LATENT VARIABLE CASE

It is well known that the graphical identifiability criteria proposed by Pearl and his colleagues are useful to evaluate total effects. However, Stanghellini (2004) pointed out that there are some situations where these identifiability criteria can not be applied to evaluate total effects. As an example, we consider the problem of evaluating the total effect $\tau_{yx}$ of $X$ on $Y$ based on the path diagram shown in Fig. 2, where $U$ is an unobserved variable and $\{X, Y, Z, W\}$ is a set of observed variables

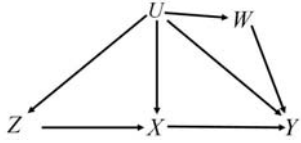

Fig. 2: Path Diagram (1)

In Fig. 2, since $U$ is an unobserved variable we can not apply the back door criterion relative to $(X, Y)$ to evaluate the total effect $\tau_{yx}$. In addition since we can not observe a set of variables that satisfies the front door criterion relative to $(X, Y)$, the front door criterion can not be applied, either. Furthermore, since there are arrows pointing from the unobserved variable $U$ to every observed variable the conditional IV method can not be applied Under such a situation, it is necessary to propose new identifiability criteria different from current results.

When we consider the linear structural equation model corresponding to the directed acyclic graph obtained by deleting from Fig. 2 an arrow pointing from $W$ to $Y$ (i.e., $\alpha_{yw} = 0$) since we can obtain the same covariance structure as the identifiable single factor model (e.g. Stanghellini, 1997), the total effect $\tau_{yx}$ is identifiable (Stanghellini, 2004) However in Fig. 2, since there is an arrow pointing from $W$ to $Y$ the number of observed covariances is less than that of the path coefficients, which indicates that the whole linear structural equation model can not be identifiable even if the variance information on $U$ is known (e.g. $\sigma_{uu} = 1$). However, the path coefficient $\alpha_{yx}$ is identifiable. This result is summarized as follows (Kuroki, 2007):

### THEOREM 1

Suppose that a set $\{X, Y, W, Z\} \cup \boldsymbol{T}$ of observed variables and an unobserved variable $U$ satisfy the following conditions in a directed acyclic graph $G$:

(1) $\{X, U\} \cup \boldsymbol{T}$ d-separates $Y$ from $Z$,

(2) $\{U\} \cup \boldsymbol{T}$ d-separates $\{X, Z\}$ from $W$, and

(3) $\{X\} \cup \boldsymbol{T}$ does not d-separate $Z$ from $W$.

When $X$ is an nondescendant of $Y$, if $\{U\} \cup \boldsymbol{T}$ satisfies the back door criterion relative to $(X, Y)$, then the total effect $\tau_{yx}$ of $X$ on $Y$ is identifiable and is given by the formula

$$\tau_{yx} = \frac{\sigma_{xw \cdot t} \sigma_{yz \cdot t} - \sigma_{zw \cdot t} \sigma_{xy \cdot t}}{\sigma_{xw \cdot t} \sigma_{zx \cdot t} - \sigma_{zw \cdot t} \sigma_{xx \cdot t}}. \quad (10)$$

□

**PROOF OF THEOREM 1**

Since $\{U\} \cup \boldsymbol{T}$ satisfies the back door criterion relative to $(X, Y)$ $\tau_{yx} = \beta_{yx \cdot ut}$ can be obtained. In addition, from Lemma 1, the following can be derived:

$$\sigma_{yz \cdot t} = \beta_{yz \cdot xut} \sigma_{zz \cdot t} + \beta_{yx \cdot uzt} \sigma_{xz \cdot t} + \beta_{yu \cdot xzt} \sigma_{uz \cdot t},$$
$$\sigma_{xy \cdot t} = \beta_{yx \cdot ut} \sigma_{xx \cdot t} + \beta_{yu \cdot xt} \sigma_{ux \cdot t},$$
$$\sigma_{xz \cdot t} = \beta_{zx \cdot tw} \sigma_{xx \cdot t} + \beta_{zw \cdot tx} \sigma_{wx \cdot t},$$
$$\sigma_{zw \cdot t} = \beta_{zw \cdot xt} \sigma_{ww \cdot t} + \beta_{zx \cdot tw} \sigma_{xw \cdot t}.$$

From condition (1) since $Y$ is conditionally independent of $Z$ given $\{X, U\} \cup \boldsymbol{T}$, $\beta_{yz \cdot xut} = 0$ can be obtained. In addition by using Lemma 2 we can obtain $\beta_{yx \cdot uzt} = \beta_{yx \cdot ut}$ and $\beta_{yu \cdot xzt} = \beta_{yu \cdot xt}$ Noting these results, we have

$$\sigma_{yz \cdot t} = \beta_{yx \cdot ut} \sigma_{xz \cdot t} + \beta_{yu \cdot xt} \sigma_{uz \cdot t}.$$

Then, we can obtain

$$\sigma_{xw \cdot t} \sigma_{yz \cdot t} - \sigma_{zw \cdot t} \sigma_{xy \cdot t}$$
$$= \beta_{yx \cdot ut} (\sigma_{xw \cdot t} \sigma_{xz \cdot t} - \sigma_{zw \cdot t} \sigma_{xx \cdot t})$$
$$+ \beta_{yu \cdot xt} (\sigma_{xw \cdot t} \sigma_{uz \cdot t} - \sigma_{zw \cdot t} \sigma_{ux \cdot t}).$$

Here since $\beta_{zw \cdot tx} \neq 0$ holds true from condition (3) and the faithful condition, from Lemma 1 and

$$\sigma_{xw \cdot t} = \beta_{xw \cdot ut} \sigma_{ww \cdot t} + \beta_{xu \cdot tw} \sigma_{uw \cdot t},$$
$$\sigma_{zw \cdot t} = \beta_{zw \cdot ut} \sigma_{ww \cdot t} + \beta_{zu \cdot tw} \sigma_{uw \cdot t},$$

we can obtain

$$\sigma_{xw \cdot t} \sigma_{xz \cdot t} - \sigma_{zw \cdot t} \sigma_{xx \cdot t}$$
$$= -\beta_{zw \cdot tx} \sigma_{ww \cdot t} (\sigma_{xx \cdot t} - \beta_{xw \cdot t} \sigma_{wx \cdot t})$$
$$= -\beta_{zw \cdot tx} \sigma_{ww \cdot t} \sigma_{xx \cdot tw} \neq 0.$$

Thus by noting that $\beta_{xw \cdot ut} = \beta_{zw \cdot ut} = 0$ can be obtained from condition (2), we have

$$\sigma_{xw \cdot t} \sigma_{uz \cdot t} - \sigma_{zw \cdot t} \sigma_{ux \cdot t} = 0.$$

By noting these results, equation (10) can be derived.
Q.E.D.

It should be noted that the assumption of the variance of an unobserved variable $U$ is not required in Theorem 1, which is different from the identification condition of factor models (e.g. Stanghellini, 1997).

In the case where there are more than one unmeasured confounder, Kuroki (2007) pointed out that the identification condition for multi-factor models (e.g.Grzebyk et al., 2004) is also useful to identify the total effects. The results can be summarized as follows.

**THEOREM 2**

Let $\boldsymbol{X} = \{X_1, \cdots, X_p\}$ be a set of observed variables and $\boldsymbol{U} = \{U_1, \cdots, U_k\}$ a set of unobserved variables in the path diagram $G$. When a linear structural equation model obtained by conditioning on $\{U_1, \cdots, U_{i-1}\}$ and marginalizing on $\{U_{i+1}, \cdots, U_k\}$ is regarded as a single factor model of $U_i$, if the single factor model of $U_i$ is identifiable for any $i(1 \le i \le k)$ $\Sigma_{xx \cdot u}$ is also identifiable.

**PROOF OF THEOREM 2**

First the covariance matrix corresponding to the linear structural equation model which is marginalized on $\{U_2, \cdots, U_k\}$ is given by

$$\Sigma_{xx} = \Sigma_{xx \cdot u_1} + \frac{1}{\sigma_{u_1 u_1}} \Sigma_{xu_1} \Sigma'_{xu_1}.$$

Then, $\Sigma_{xx \cdot u_1}$ is identifiable from the assumption

Here, we assume that $\Sigma_{xx \cdot u_1 \cdots u_{i-1}}$ is identifiable for $i(\ge 2)$. Then, the covariance matrix corresponding to the linear structural equation model which is marginalized on $\{U_{i+1}, \cdots, U_k\}$ and conditioned on $\{U_1, \cdots, U_{i-1}\}$ is given by

$\Sigma_{xx \cdot u_1 \cdots u_{i-1}}$
$= \Sigma_{xx \cdot u_1 \cdots u_i} + \frac{1}{\sigma_{u_i u_i \cdot u_1 \cdots u_{i-1}}} \Sigma_{xu_i \cdot u_1 \cdots u_{i-1}} \Sigma'_{xu_i \cdot u_1 \cdots u_{i-1}}.$

Thus, $\Sigma_{xx \cdot u_1 \cdots u_i}$ is identifiable from the assumption By repeating this procedure, the result can be obtained. $QED$

When Theorem 2 holds true for $\{X, Y\} \cup \boldsymbol{Z} \subset \boldsymbol{X}$ if $\boldsymbol{Z} \cup \boldsymbol{U}$ satisfies the back door criterion relative to $(X, Y)$ the total effect $\tau_{yx}$ is identifiable

As an example, we consider the problem of evaluating total effect $\tau_{yx}$ in the path diagram shown in Fig. 3. Although we can not apply the identifiability criteria proposed by Pearl and his colleagues, since Theorem 2 holds true, the total effect $\tau_{yx}$ is identifiable and is given by the formula

$$\tau_{yx} = \frac{\sigma_{yz_1} \sigma_{z_2 w_1} - \sigma_{yw_1} \sigma_{z_1 z_2}}{\sigma_{xz_1} \sigma_{z_2 w_1} - \sigma_{xw_1} \sigma_{z_1 z_2}}.$$

It is interesting that the above equation does not include the covariance parameter about $W_2$.

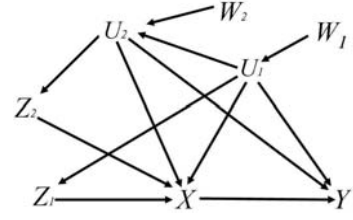

Fig. 3: Path Diagram (2)

### 3.4 IDENTIFAIBLITY CRITERION: SELECTION BIAS CASE

Selection bias is another case that the identifiability criteria proposed by Pearl and his colleagues can not be applied to evaluate total effects. Consider the identification problem for the total effect $\tau_{yx}$ based on the path diagram shown in Fig. 4, which indicates that sample selection is conducted according to a criterion $a \le S \le b$. Then, $S$ is called a selection variable. In addition, $\{X, Y, Z, W\}$ is a set of observed variables.

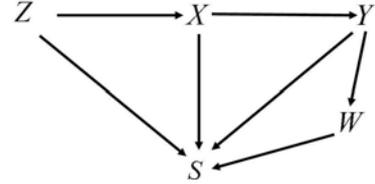

Fig. 4: Path Diagram (3)

In Fig. 4, since a sample is selected from the population using such a criterion as $a \le S \le b$, the statistical dependencies among $\{X, Y, Z, W\}$ are biased. Thus, we can not apply any identifiability criteria proposed by Pearl and his colleagues to identify the total effect. On the other hand, when we consider the linear structural equation model corresponding to the directed acyclic graph obtained by deleting from Fig. 4 an arrow pointing from $Y$ to $W$ (i.e. $\alpha_{wy} = 0$), since the number of the observed covariances is equal to that of the path coefficients, the total effect $\tau_{yx}$ can be evaluated through the observed covariances. However, in Fig. 4, since the number of the observed covariances is less than that of the path coefficients, the whole linear structural equation model is not identifiable. But, the total effect $\tau_{yx}$ is identifiable through the following theorem.

**THEOREM 3**

Suppose a set $\{X, Y, W, Z\} \cup \boldsymbol{T}$ of observed variables and a selection variable $S$ satisfy the following conditions in a directed acyclic graph $G$:

(1) $\{X\} \cup \boldsymbol{T}$ d-separates $Y$ from $Z$,

(2) $\boldsymbol{T}$ d-separates $\{X, Z\}$ from $\{W\}$,

(3) $\{X\} \cup \boldsymbol{T}$ does not d-separate $S$ from $Z$, and

(4) $\boldsymbol{T}$ does not d-separate $S$ from $W$.

When $X$ is a nondescendant of $Y$, if $\boldsymbol{T}$ satisfies the back door criterion relative to $(X,Y)$, then the total effect $\tau_{yx}$ of $X$ on $Y$ is identifiable and is given by the formula

$$\tau_{yx} = \frac{\sigma_{xw \cdot ts^*}\sigma_{yz \cdot ts^*} - \sigma_{zw \cdot ts^*}\sigma_{xy \cdot ts^*}}{\sigma_{xw \cdot ts^*}\sigma_{zx \cdot ts^*} - \sigma_{zw \cdot ts^*}\sigma_{xx \cdot ts^*}}. \quad (11)$$

□

**PROOF OF THEOREM 3**

Since $\boldsymbol{T}$ satisfies the back door criterion relative to $(X,Y)$, $\tau_{yx} = \beta_{yx \cdot t}$ can be obtained. In addition, from Lemma 1 and condition (2),

$$\begin{aligned}
\sigma_{yz \cdot ts^*} &= \sigma_{yz \cdot xt} + \beta_{yx \cdot tz}\sigma_{xz \cdot t} - \beta_{ys \cdot t}\beta_{zs \cdot t}\check{\sigma}_{ss \cdot t} \\
\sigma_{xy \cdot ts^*} &= \sigma_{yx \cdot t} - \beta_{xs \cdot t}\beta_{ys \cdot t}\check{\sigma}_{ss \cdot t}, \\
\sigma_{zx \cdot ts^*} &= \sigma_{zx \cdot t} - \beta_{zs \cdot t}\beta_{xs \cdot t}\check{\sigma}_{ss \cdot t}, \\
\sigma_{zw \cdot ts^*} &= -\beta_{zs \cdot t}\beta_{ws \cdot t}\check{\sigma}_{ss \cdot t}, \\
\sigma_{xw \cdot ts^*} &= -\beta_{xs \cdot t}\beta_{ws \cdot t}\check{\sigma}_{ss \cdot t}.
\end{aligned}$$

From conditions (1), since $Y$ is conditionally independent of $Z$ given $\{X\}\cup\boldsymbol{T}$, $\sigma_{yz \cdot xt} = 0$ can be obtained. In addition, from Lemma 2, we can obtain $\beta_{yx \cdot zt} = \beta_{yx \cdot t}$. Using these results, we have

$$\sigma_{yz \cdot ts^*} = \beta_{yx \cdot t}\sigma_{xz \cdot t} - \beta_{ys \cdot t}\beta_{zs \cdot t}\check{\sigma}_{ss \cdot t}.$$

Thus, we can obtain

$$\begin{aligned}
&\sigma_{xw \cdot ts^*}\sigma_{yz \cdot ts^*} - \sigma_{zw \cdot ts^*}\sigma_{xy \cdot t}s^* \\
&= \beta_{yx \cdot t}\beta_{ws \cdot t}\check{\sigma}_{ss \cdot t}\sigma_{xx \cdot t}(\beta_{xs \cdot t}\beta_{zx \cdot t} - \beta_{zs \cdot t}) \\
&= -\beta_{yx \cdot t}\beta_{ws \cdot t}\check{\sigma}_{ss \cdot t}\sigma_{xx \cdot t}\sigma_{zs \cdot xt}/\sigma_{ss \cdot t}.
\end{aligned}$$

Here, since $\beta_{ws \cdot t}\beta_{zs \cdot xt}\neq 0$ holds true from conditions (3) and (4) and the faithful condition, according to Lemma 1, we can obtain

$$\begin{aligned}
&\sigma_{xw \cdot ts^*}\sigma_{xz \cdot ts^*} - \sigma_{zw \cdot ts^*}\sigma_{xx \cdot ts^*} \\
&= -\beta_{ws \cdot t}\sigma_{xx \cdot t}\check{\sigma}_{ss \cdot t}(\beta_{zs \cdot t} - \beta_{xs \cdot t}\beta_{zx \cdot t}) \\
&= -\beta_{ws \cdot t}\check{\sigma}_{ss \cdot t}\sigma_{xx \cdot t}\sigma_{zs \cdot xt}/\sigma_{ss \cdot t}.
\end{aligned}$$

By noting these results, equation (11) is derived.
Q.E.D.

### 3.5 IDENTIFAIBLITY CRITERION: LATENT VARIABLE AND SELECTION BIAS CASE

Finally, we consider the case where both latent variables $\boldsymbol{U}$ and selection bias according to a selection criterion $a \leq S \leq b$ exist. Let $\boldsymbol{X}$ ($X, Y \in \boldsymbol{X}$) and $\boldsymbol{T}$ be sets of observed variables, the steps for judging whether or not the total effect is identifiable are as follows:

**Step 1**: Check whether or not the combination of a subset of $\boldsymbol{X}\setminus\{X,Y\}$ and $\boldsymbol{U}\cup\boldsymbol{T}$ satisfies the back door criterion relative to $(X,Y)$. If the answer is affirmative, go to Step 2.

**Step 2**: By noting that

$$\Sigma_{xx \cdot ts^*} = \Sigma_{xx \cdot t} - B_{xs \cdot t}B'_{xs \cdot t}\check{\sigma}_{ss \cdot t},$$

check whether or not the structure of zeros in $\Sigma_{xx \cdot t}$ (or $\Sigma^{-1}_{xx \cdot t}$) is such that every connectivity component of the complementary graph of $G^{x \cdot t}_{cov}$ (or $G^{x \cdot t}_{con}$) contains an odd cycle. If the answer is affirmative, since $\Sigma_{xx \cdot t}$ is identifiable (Stanghellini and Wermuth, 2005), then go to Step 3.

**Step 3**: Check whether Theorem 2 holds true for $\Sigma_{xx \cdot t}$ with regard to $\boldsymbol{U}$. If the answer is affirmative, since $\Sigma_{xx \cdot tu}$ is identifiable, then we can evaluate the total effect $\tau_{yx}$ of $X$ on $Y$.

## 4 APPLICATION

The above results are applicable to analyze the data from a study about setting up painting conditions of car bodies, reported by Okuno et al. (1986). The data was collected with the purpose of setting up the process conditions, in order to increase transfer efficiency. The size of the sample is 38 and the variables of interest, each of which has zero mean and variance one, are the following:

Painting Condition  Dilution Ratio ($X_1$),
  Degree of Viscosity ($X_2$), Painting Temperature ($X_8$)

Spraying Condition  Gun Speed ($X_3$),
  Spray Distance ($X_4$), Atomizing Air Pressure ($X_5$),
  Pattern Width ($X_6$), Fluid Output ($X_7$)

Environment Condition  Temperature ($X_9$),
  Degree of Moisture ($X_{10}$)

Response: Transfer Efficiency ($Y$)

Concerning this process, Kuroki et al. (2003) presented the path diagram shown in Fig. 5 (for the detail, see Kuroki et al., 2003). Based on the path diagram, Kuroki et al. (2003) presented the estimated correlation matrix. We here provide a part of the correlation matrix in Table 1. From Table 1, we assume that the covariance information on $X_4$ and $X_{10}$ is not obtained.

Although $X_1, \cdots, X_6$ are considered to be controllable variables in Okuno et al. (1986), $X_2$ and $X_6$ are taken as treatment variables from controllable variables in order to evaluate their total effects from nonexperimental data in this paper.

Table 2 shows the selected variables for estimating total effects. The treatment variables of interest are listed in the first column. The second column shows

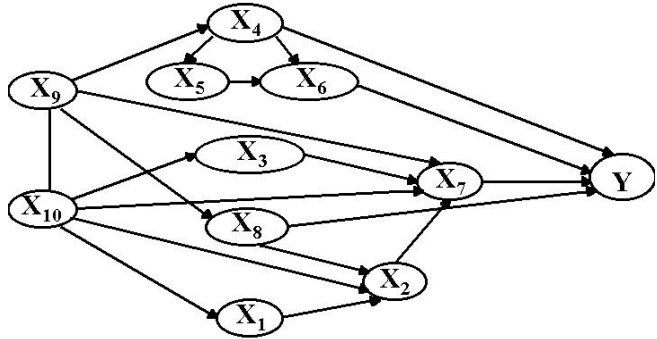

Fig. 5   Path Diagram (Kuroki et al., 2003)

Table 1   Estimated Correlation Matrix (Kuroki et al., 2003)

|       | $X_1$  | $X_2$  | $X_5$  | $X_6$  | $X_8$  | $X_9$  | $Y$    |
|-------|--------|--------|--------|--------|--------|--------|--------|
| $X_1$ | 1.000  | −0.736 | 0.028  | −0.042 | 0.216  | 0.283  | −0.091 |
| $X_2$ | −0.736 | 1.000  | −0.063 | 0.095  | −0.684 | −0.635 | 0.326  |
| $X_5$ | 0.028  | −0.063 | 1.000  | 0.291  | 0.076  | 0.099  | −0.277 |
| $X_6$ | −0.042 | 0.095  | 0.291  | 1.000  | −0.114 | −0.149 | −0.250 |
| $X_8$ | 0.216  | −0.684 | 0.076  | −0.114 | 1.000  | 0.761  | −0.493 |
| $X_9$ | 0.283  | −0.635 | 0.099  | −0.149 | 0.761  | 1.000  | −0.475 |
| $Y$   | −0.091 | 0.326  | −0.277 | −0.250 | −0.493 | −0.475 | 1.000  |

Table 2   Estimates of Total Effects

| treatment | covariates | total effect |
|-----------|------------|--------------|
| $X_2$ | $Z = X_1, W = X_9, T = X_8$ | -0.116 |
| $X_6$ | $Z = X_5, W = X_9, T = \phi$ | -0.465 |

sets of covariates used for identifying total effects. The third columns shows the estimates of total effects. First, consider a situation that we wish to evaluate the total effect of $X_2$ on $Y$. Then, it can be recognized that the total effect can not be evaluated based on the back door criterion or the conditional IV method, because the covariance information on $X_{10}$ can not be obtained from Table 2. In addition, since $X_{10}$ exists in the back door path between $X_2$ and $X_7$, the front door criterion can not be applied, either  However since a set of variables provided in the second column satisfies the conditions in Theorem 1, the total effect can be evaluated by using equation (10)

Next, consider a situation that we wish to evaluate the total effect of $X_6$ on $Y$. Then, it can be recognized that the total effect can not be evaluated based on the back door criterion or the conditional IV method, because the covariance information on $X_4$ can not be obtained from Table 2. In addition, there is not a set of variables satisfying the front door criterion  However  since a set of variables provided in the second column satisfies the conditions in Theorem 1, the total effect can be evaluated by using equation (10)

## 5   DISCUSSION

This paper discussed identification problems for total effects based on causal modeling in observational studies with latent variables and selection bias. In order to derive the graphical identifiability criteria, we introduced identification condition for factor models to the identification problem of total effects. In addition, we pointed out that there are some cases where the total effect is identifiable even when the identification condition for factor models does not hold true. Furthermore, we proposed new identification conditions of total effects, and provided the closed form expression of the identifiable total effects. The results of this paper help us judge from graph structure whether the total effect can be evaluated from observational studies in the presence of latent variables and selection bias.

### ACKNOWLEGDEMENT

This research was supported by the Ministry of Education, Culture, Sports, Science and Technology of Japan, the Kurata Foundation and the Mazda Foundation.

### REFERENCES

Bowden, R. J. , and Turkington, D. A. (1984). *Instrumental Variables*, Cambridge University Press.

Brito, C. and Pearl, J. (2002). Generalized instrumental variables. *Proceeding of the 18th Conference on Uncertainty in Artificial Intelligence*, 85-93.

Cochran, W.G. (1938). The omission or addition of an independent variate in multiple linear regression, *Sup-*

*plement to the Journal of the Royal Statistical Society*, **5**, 171-176.

Cooper, G. F. (2000). A Bayesian method for causal modeling and discovery under selection. *Proceedings of the Conference on Uncertainty in Artificial Intelligence*, **16**, 98-106.

Greenland, S. (2003). Quantifying biases in causal models: Classical confounding versus collider-stratification bias. *Epidemiology*, **14**, 300-306.

Grzebyk, M., Wild, P. and Chouaniere, D. (2004). On identification of multi-factor models with correlated residuals. *Biometrika*, **91**, 141-151.

Hernan, M. A. , Hernandez-Diaz, S. and Robins, J. M. (2004). A structural approach to selection bias. *Epidemiology*, **15**, 615-625.

Huang, Y. and Valtorta, M. (2006). Pearl's Calculus of Intervention is Complete. *Proceedings of the Conference on Uncertainty in Artificial Intelligence*, **22**, 437-444.

Johnson, N. L. & Kotz, S. (1972). *Distributions in Statistics : Continuous Multivariate Distributions*. New York: John Wiley & Sons.

Kuroki, M.(2007). Identifiability Criteria for Total Effects in the Presence of Unmeasured Confounders (In Japanese),*Japanese Journal of Applied Statistics*,**36**, 71-85.

Kuroki, M. and Cai, Z. (2006). On Recovering Population's Covariance Matrix in the Presence of Selection Bias, *Biometrika*, **93**, 601-611.

Kuroki, M., Miyakawa, M. and Cai, Z. (2003). Joint Causal Effect in Linear Structural Equation Model and Its Application to Process Analysis, *Proceedings of the Workshop on Artificial Intelligence and Statistics*, **9**, 70-77

Lauritzen, S. L. (1996). *Graphical models*, Clrendon Press Oxford.

Okuno, T., Katayama, Z., Kamigori, N., Itoh, T., Irikura, N. and Fujiwara, N. (1986). *Kougyou ni okeru Tahenryou Data no Kaiseki* (In Japanese), Nikkagiren, Tokyo.

Pearl, J. (2000). *Causality: Models, reasoning, and inference*. Cambridge University Press.

Rao, C. R. (1973). *Linear statistical inference and its applications*. John Wiley & Sons.

Richardson, T. S. and Spirtes, P. (2002). Ancestral graph markov models. *Annals of Statistics*, **30**, 962-1030.

Shpitser,I. and Pearl, J. (2006). Identification of Joint Interventional Distributions in Recursive Semi-Markovian Causal Models. *Proceedings of the National Conference on Artificial Intelligence*,**21**, 1219-1226.

Spirtes, P., Glymour, C. and Scheines, R. (2000). *Causation, prediction, and search, 2nd edition*, MIT Press.

Spirtes, P. , Meek, C. , and Richardson, T. (1999). An algorithm for causal inference in the presence of latent variables and selection bias. In Glymour, C. and Cooper, G. , editors, *Computation, Causation, and Discovery*, 211-252.

Stanghellini, E. (1997). Identification of a single-factor models using graphical Gaussian rules. *Biometrika*, **84**, 241-244.

Stanghellini, E (2004) Instrumental variables in Gaussian directed acyclic graph models with an unobserved confounder *Environmetrics*, **15**, 463-469

Stanghellini, E. and Wermuth, N. (2005). On the identification of directed acyclic graph models with one hidden variable. *Biometrika*, **92**, 337-350.

Tian, J. (2004). Identifying linear causal effects. *Proceeding of the Nineteenth National Conference on Artificial Intelligence*, 104-111.

Wermuth, N. (1989). Moderating effects in multivariate normal distributions. *Methodika*, **3**, 74-93.

Whittaker, J. (1990). *Graphical Models in Applied Multivariate Statistics*, John Wiley and Sons.